# Identifying and modeling patterns of tetrapod vertebrate mortality rates in the Gulf of Mexico oil spill

F. J. Antonio[a,b,*], R. S. Mendes[a,b] & S. M. Thomaz[c]

[a] Maringá State University, Department of Physics, Av. Colombo 5790, Maringá, 87010-900, Brazil.
[b] National Institute of Science and Technology for Complex Systems, CNPq, Rua Xavier Sigaud 150, 22290-180, Rio de Janeiro, RJ, Brazil.
[c] Maringá State University, Department of Biology – Nupélia, Av. Colombo 5790, Maringá, 87010-900, Brazil.

**Abstract**
The accidental oil spill in the Gulf of Mexico in 2010 has caused perceptible damage to marine and freshwater ecosystems. The large quantity of oil leaking at a constant rate and the long duration of the event caused an exponentially increasing mortality of vertebrates. Using data provided by NOAA and USFWS, we assessed the effects of this event on birds, sea turtles, and mammals. Mortality rates (measured as the number of carcasses recorded per day) were exponential for all three groups. Birds were the most affected group, as indicated by the steepest increase of mortality rates over time. For sea turtles and mammals, an exponential increase in mortality was observed after an initial delay. These exponential behaviors are consistent with a unified scenario for the mortality rate for tetrapod vertebrates. However, at least for mammals, pre-spill data seem to indicate that the growth in the mortality rate is not entirely a consequence of the spill.

Oil spills damage both marine and freshwater ecosystems. Among the best-reported spills are the Exxon Valdez in Alaska (Bohannon et al., 2002) and the Prestige in Spain (Castege et al., 2007). These and other oil spills are limited in space, time, and volume of oil leaked, compared to the Gulf of Mexico Deepwater Horizon accident (Estes, 1991; Marine Conservation Society, 2010).

The most visible and immediate impact of oil on wildlife is its adhesion to organisms, especially after the sludge washes ashore. However, other dangerous impacts are related to mutagenic and/or carcinogenic polycyclic aromatic hydrocarbons present in the oil (Samanta et al., 2002) and sub-lethal effects such as physiological or endocrine disruptions (Bilbao et al., 2010). The Gulf of Mexico spill offers an excellent opportunity for ecologists, toxicologists, and environmental physiologists to understand the effects of oil on the biota and to identify the responses of the organisms affected by the spill.

We used the estimated volume of oil leaked and a consolidated number of dead animals provided by the National Oceanic and Atmospheric Administration (NOAA) and the U.S. Fish and Wildlife Service (USFWS) available on the website www.fws.gov/home/dhoilspill/collectionreports.html to accomplish two objectives: *(i)* to identify patterns and model the response of tetrapod vertebrates to this extreme oil release, and *(ii)* to assess which group(s) of tetrapod vertebrates (birds, sea turtles, and mammals) was most affected by the spill. Our results will help to predict the outcome of impacts from oil spills, and also to evaluate which groups of animals should be prioritized in rescues during similar events.

Carcasses were collected along *ca.* 2,600 km, in the US coast (Gulf of Mexico). The numbers of animals collected were reported to the Unified Area Command by the USFWS, NOAA, area incident commands, rehabilitation centers, and other authorized sources operating within the Deepwater Horizon/BP (British Petroleum) incident impact area. Because the results are based only on rescues accomplished by these agencies, we assumed a constant sampling effort. We also assumed that even if effort was not completely constant, the data are adequate for accomplishing at least our second objective (to assess which group of vertebrates is most sensitive to oil), because over a given period, the effort was the same for each group of animals. In addition, our models are independent of the number of individuals in a population when the number of deaths is very small relative to population sizes (this assumption will be argued later in the text).

---

* Corresponding author. Tel./fax: +55 44 32634623.
E-mail address: fern_jose@yahoo.com.br (F. J. Antonio).

The oil spill lasted from 22 April to 16 July 2010. The volume of oil leaked (P) accumulated at a constant rate ($R_P$ = dP/dt) (Fig. 1(A)). One should note that the NOAA and USFWS dataset does not contain measurements related to the forces moving the oil spill or dispersants nor the coordinates where the carcasses were found. Given these characteristics, a simplified model of the impacts on wildlife can be generated.

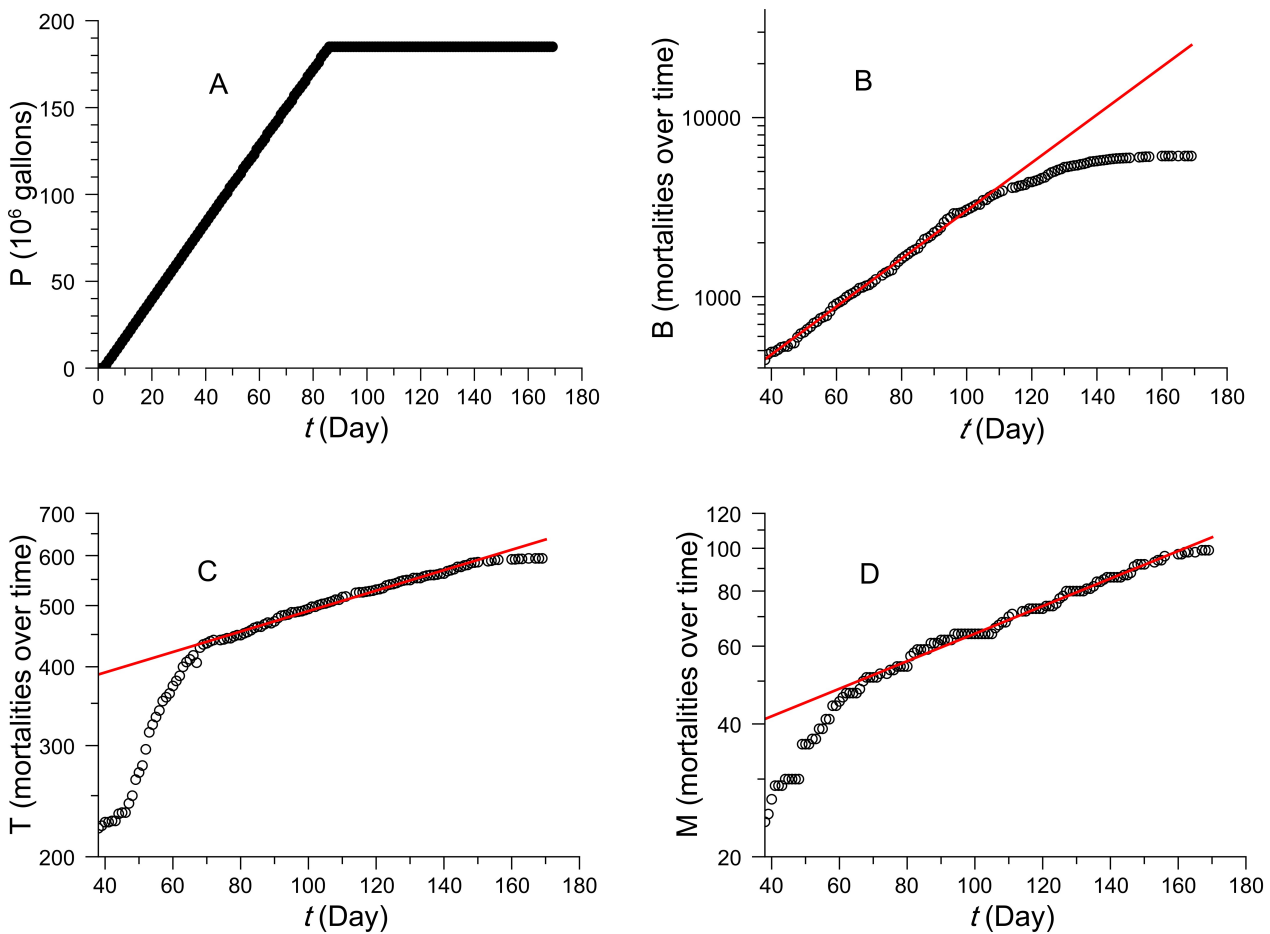

Fig. 1: Cumulative daily increase in estimated oil spilled/P (A), and cumulative recorded numbers of carcasses of birds/B (B), sea turtles/T (C), and mammals/M (D). The solid lines are exponential functions adjusted to the data.

As a consequence of the oil spill, there was increasing mortality of vertebrates, whose carcasses started to be reported after the 38[th] day (May 28). In our analysis we considered the first 140 days reporting period. The highest impact during this period was recorded for birds, with an exponential increase in the cumulative number of carcasses (B; Fig. 1(B)). Consequently, the daily mortality rate ($R_B$ = dB/dt) was also an exponential function. The intrinsic increase of mortality rates for this group was $r_B = 0.01338 \pm 0.00008$ d$^{-1}$. A smaller number of sea turtle carcasses (T) was recorded (Fig. 1(C)). There was an exponential increase in mortality for this group, also leading to exponential daily mortality rates ($R_T = dT/dt$). However, we identified an initial delay before the establishment of the exponential increase (*ca.* day 68/June 27). An even smaller number of mammal carcasses (M) was recorded (Fig. 1(D)). For mammals, there was also an initial delay before the establishment of an exponential increase (*ca.* day 77/July 6). Although not as clear as that observed for birds and turtles, the data obtained for mammals suggest an exponential increase in mortality and an exponential daily mortality rate ($R_M = dM/dt$). The intrinsic increase of mortality rates for turtles ($r_T$) was $0.00162 \pm 0.00002$ d$^{-1}$ and for mammals ($r_M$) $0.00312 \pm 0.00003$ d$^{-1}$. The lack of an initial delay for birds (Fig. 1(B)), as observed for these last two groups, may be associated with the fact that birds are more sensitive and arrive at the beaches long before the oil reaches the shore (Castege et al., 2007). However, birds were the first group to recover from the oil spill, as indicated by their mortality rates which decreased after the 97$^{th}$ day (July 26), whilst the mortality of turtles and mammals was still rising exponentially until the 155$^{th}$ day (September 22) and the 163$^{th}$ day (September 30) respectively. These exponential regularities can be expressed mathematically by the equation $dR_i/dt = r_i R_i$ with i =B, T and M.

In our modeling of vertebrate mortality rates, two other desired aspects could be taken into account: *(i)* mortality rates $R_i$ should be constant when the oil leak stopped ($R_P = 0$), and *(ii)* the intrinsic increase in the mortality rate of each group of organisms should depend on its sensitivity to petroleum. These aspects and the reported exponential behaviors are naturally accomplished if the intrinsic increases of mortality rates are given by $r_i = a_i R_P$, with $a_i$ being related only to the petroleum sensitivity of the i$^{th}$ group of organisms. In fact, the mortality rate ($R_i$) for all groups becomes constant ($dR_i/dt = 0$) when $R_P = 0$.

Note that when one considers population deaths it is desirable to take the number of deaths into account in a time interval for a given population size. This aspect can be considered via the relative mortality rate $\check{R}_i = R_i/P_i$ where $P_i$ is the population size of the i$^{th}$ group. In most cases, as occurred in our work, only data about the number of deaths along time, but not the population size, are available. However, it is desirable to identify intrinsic patterns of the system, despite the lack of more complete data about the population sizes. In our investigation we identified patterns when the number of carcasses is much lower than the population size. In fact, if $P_i$ is much higher than the number of deaths (more precisely, $R_i \ll r_i P_i$) and therefore almost constant along the Gulf accident, $d\check{R}_i/dt$ is very well approximated by $1/P_i \, dR_i/dt$. Consequently, $r_i$ ($a_i$) is an intrinsic parameter for each group which does not depend on the population size since $r_i = 1/R_i \, dR_i/dt = 1/\check{R}_i \, d\check{R}_i/dt$. Thus, the parameter $r_i$ ($a_i$) is appropriate to compare different groups of animals, even without knowing their population sizes, and we interpret higher $r_i$ ($a_i$) as higher sensitivity.

To support this assumption of effective independence of the population size, we take as an example the number of carcasses for mammals (*ca.* 100), which represents less than 0.15% of the population of dolphins in the Gulf (*ca.* 75,000; http://www.telegraph.co.uk/earth/wildlife/7713466/Gulf-of-Mexico-oil-spill-dead-dolphins-found-washed-up-on-US-coast.html). Note that this rate is even lower if we consider that the real number of mammals is higher because it includes several other species, supporting the use of $r_i$ ($a_i$) to compare different groups of animals.

A concern with our data is the limited access to the data about mortality rates in a context of pre-spill mortality and thus we do not know how much of the measured mortality is associated with the spill, compared to other sources of mortality. Although we did not get data for all groups, we obtained the number of carcasses for turtles (http://www.sefsc.noaa.gov/stssnrep/SeaTurtleReportI.do?action=reportquery; accessed in January 2011) and stranded cetaceans (http://www.nmfs.noaa.go

v/pr/health/mmume/cetacean_gulfofmexico2010.htm; accessed in May 2011) before the pollution event, and we could compare these data with our data. According to our expectations, the mortality rates for turtles were within the same order of magnitude before spill (~3 $d^{-1}$) and after spill cessation (~2 $d^{-1}$). Thus, it is also reasonable to assume that the much higher mortality rates obtained during the spill (~8 $d^{-1}$ between days 38 (May 28) and 68 (June 28)) are associated with the oil impacts. However, for cetaceans (a fraction of mammals) a different figure was found. For example, in March 2010 (*i.e.*, before the event) the number of stranded cetaceans was 3.4 times higher than the historical average recorded in this period. Thus, considering cetaceans as a good approach for mammals, the increase in the number of carcasses found does not seem to be entirely associated to the spill event.

Our model could be improved to take the initial delay into account. To achieve this improvement, a time delay equation (Murray, 2002) $dR_i(t)/dt = r_i R_i(t-\tau_i)$ may be employed, where $\tau_i$ is a suitable transient time and i = B, T and M. A more detailed model accomplishes the interpolating region between the exponential and the asymptotic regimes. The exponential behavior recorded for the mortality of animals during the Gulf of Mexico oil spill and the equations describing this particular scenario could be tested in accidents involving other pollutants.

**Acknowledgements**
We thank the Brazilian agencies CNPq and CAPES for financial support in the form of a PhD scholarship to FJ Antonio and Productivity Research grants to RS Mendes and SM Thomaz.

**References**

Bilbao, E., Raingeard, D., de Cério, O. D., Ortiz-Zarragoitia, M., Ruiz, P., Izagirre, U., Orbea, A., Marigomez, I., Cajaraville, M. P., Cancio, I. 2010. Effects of exposure to Prestige-like heavy fuel oil and to perfluorooctane sulfonate on conventional biomarkers and target gene transcription in the thicklip grey mullet *Chelon labrosus*. Aquat. Toxicol. 98, 282-296.
Bohannon, J., Bosch, X., Withgott, J., 2002. Scientists brace for bad tidings after spill. Science 298, 296-297.
Castege, I., Lalanne, Y., Gouriou, V., Hemery, G., Girin, M., D'Amico, F., Mouches, C., D'Elbee, J., Soulier, L., Pensu, J., Lafitte, D., Pautrizel, F., 2007. Estimating actual seabirds mortality at sea and relationship with oil spills: lesson from the "Prestige" oil spill in Aquitaine (France). Ardeola 54, 289-307.
Estes, J.A., 1991. Catastrophes and conservation: lessons from sea otters and the Exxon Valdez. Science. 254, 1596.
Marine Conservation Society, 2010. Deepwater Horizon oil spill: US prepares for the worst. Mar. Pollut. Bull. 60, 793.
Murray, J.D., Mathematical Biology: I. An Introduction, third ed., Springer, 2002.
Samanta, S.K., Singh, O.V., Jain, R.K., 2002. Polycyclic aromatic hydrocarbons: environmental pollution and bioremediation. Trends Biotechnol. 20, 243-248.